\documentclass[preprint,twoside]{aastex}
\usepackage{graphicx}

\newcommand{\ms}{\mbox{\rm ms} }

\newcommand{\mHz}{\mbox{\rm mHz}}
\newcommand{\Hz}{\mbox{\rm Hz} }

\newcommand{\GHz}{\mbox{\rm GHz} }

\newcommand{\mkelvin}{\mbox{\rm mK} }
\newcommand{\ukelvin}{\mbox{$\mu$\rm K}}

\newcommand{\ukelvinrms}{\mbox{$\mu$\rm K rms}}

\newcommand{\wmap}{\mbox{\sl WMAP}}
\newcommand{\Barnes}{{C. Barnes}}
\newcommand{\Bennett}{{C. L. Bennett}}

\newcommand{\Halpern}{{M. Halpern}}

\newcommand{\Hinshaw}{{G. Hinshaw}}
\newcommand{\Jarosik}{{N. Jarosik}}
\newcommand{\Kogut}{{A. Kogut}}

\newcommand{\Limon}{{M. Limon}}
\newcommand{\Meyer}{{S. S. Meyer}}

\newcommand{\Page}{{L. Page}}

\newcommand{\Spergel}{{D. N. Spergel}}
\newcommand{\Tucker}{{G. S. Tucker}}

\newcommand{\Weiland}{{J. L. Weiland}}
\newcommand{\Wollack}{{E. Wollack}}
\newcommand{\Wright}{{E. L. Wright}}

\newcommand{\Brown}{{Dept. of Physics, Brown University, %
            Providence, RI 02912}}
\newcommand{\Goddard}{{Code 685, Goddard Space Flight Center, %
            Greenbelt, MD 20771}}
\newcommand{\NRCFellow}{{National Research Council (NRC) Fellow}}
\newcommand{\PrincetonPhysics}{{Dept. of Physics, Jadwin Hall, %
            Princeton, NJ 08544}}
\newcommand{\PrincetonAstro}{{Dept of Astrophysical Sciences, %
            Princeton University, Princeton, NJ 08544}}
\newcommand{\SSAI}{{Science Systems and Applications, Inc. (SSAI), %
            10210 Greenbelt Road, Suite 600 Lanham, Maryland 20706}}
\newcommand{\UBC}{{Dept. of Physics and Astronomy, University of %
            British Columbia, Vancouver, BC  Canada V6T 1Z1}}
\newcommand{\UChicago}{{Depts. of Astrophysics and Physics, EFI and CfCP, %
            University of Chicago, Chicago, IL 60637}}
\newcommand{\UCLA}{{UCLA Astronomy, PO Box 951562, Los Angeles, CA 90095-1562}}

\newcommand{\sinc}{{\rm sinc}}

\begin{document}

\title{First Year {\sl Wilkinson Microwave Anisotropy Probe} (\wmap) Observations: On-Orbit Radiometer Characterization}
\author{\Jarosik \altaffilmark{1}, 
\Barnes  \altaffilmark{1}, 
\Bennett \altaffilmark{2},
\Halpern \altaffilmark{3},
\Hinshaw \altaffilmark{2}, 
\Kogut \altaffilmark{2}, 
\Limon \altaffilmark{2,4}, 
\Meyer \altaffilmark{5},
\Page \altaffilmark{1},
\Spergel \altaffilmark{6},
\Tucker \altaffilmark{7,2,4},
\Weiland \altaffilmark{8},
\Wollack \altaffilmark{2},
\Wright \altaffilmark{9}}

\altaffiltext{1}{\PrincetonPhysics}
\altaffiltext{2}{\Goddard}
\altaffiltext{3}{\UBC}
\altaffiltext{4}{\NRCFellow}
\altaffiltext{5}{\UChicago}
\altaffiltext{6}{\PrincetonAstro}
\altaffiltext{7}{\Brown}
\altaffiltext{8}{\SSAI}
\altaffiltext{9}{\UCLA}
\altaffiltext{10}{\wmap~is the result of a partnership between Princeton 
                 University and NASA's Goddard Space Flight Center. The HEMT amplifiers
used in \wmap~were supplied by the National Radio Astronomy Observatory. Scientific 
		 guidance is provided by the \wmap~Science Team.}
\email{jarosik@pupgg.princeton.edu}

\begin{abstract}
The \wmap~satellite has completed one year of measurements of the Cosmic Microwave Background (CMB)
radiation using 20 differential high-electron-mobility-transistor (HEMT) based  radiometers.
All the radiometers are functioning nominally, and characterizations of the 
on-orbit radiometer performance are presented, with an emphasis on properties that are required for the 
production of sky maps from the time ordered data. A radiometer gain model, used to
smooth and interpolate the CMB dipole gain  measurements  is also presented. No degradation in the
sensitivity of any of the radiometers has been observed during the first year of observations.
\end{abstract}


\section{INTRODUCTION}
The {\sl Wilkinson Microwave Anisotropy Probe} (\wmap), launched 2001 June 30,  is a
Medium-class Explorer (MIDEX) mission designed to produce full sky maps
of the cosmic microwave background (CMB) radiation. The \wmap~instrument
is comprised of a dual set  off-axis Gregorian telescopes \citep{page/etal:2003} coupled to 20 differential
radiometers \citep{jarosik/etal:2003}. \wmap~scans the sky from near the second  Earth-Sun Lagrange
point with a 129 s spin period and a 1 hour precession period
 \citep{bennett/etal:2003}.  The resulting scan pattern is highly interconnected, 
 minimizing systematic errors in the reconstructed maps.
 \wmap~utilizes
high electron mobility transistor (HEMT) based amplifiers \citep{pospieszalski/etal:2000} with cooled input stages
to reduce the noise of the radiometers. This paper describes the on-orbit performance
of the radiometers with emphasis on those characteristics which  determine the
quality of the sky maps. Design details and terminology used for the optics, radiometers and
overall mission can be found in \citet{page/etal:2003}, 
\citet{jarosik/etal:2003} and \citet{bennett/etal:2003}
respectively.

Detailed knowledge of the performance of the \wmap~optics and
radiometers is essential in order to convert the raw time ordered data (TOD) into
accurate maps of the microwave sky. \citep{hinshaw/etal:2003b} Characterization of both the optics and radiometers
employ a combination of pre-launch ground based measurements and on-orbit data. 
Characterizations of the main beam and sidelobe response of the optics are presented
in   \citet{page/etal:2003b} and \citet{barnes/etal:2003}.    

\begin{figure*}
\includegraphics[width=6in]{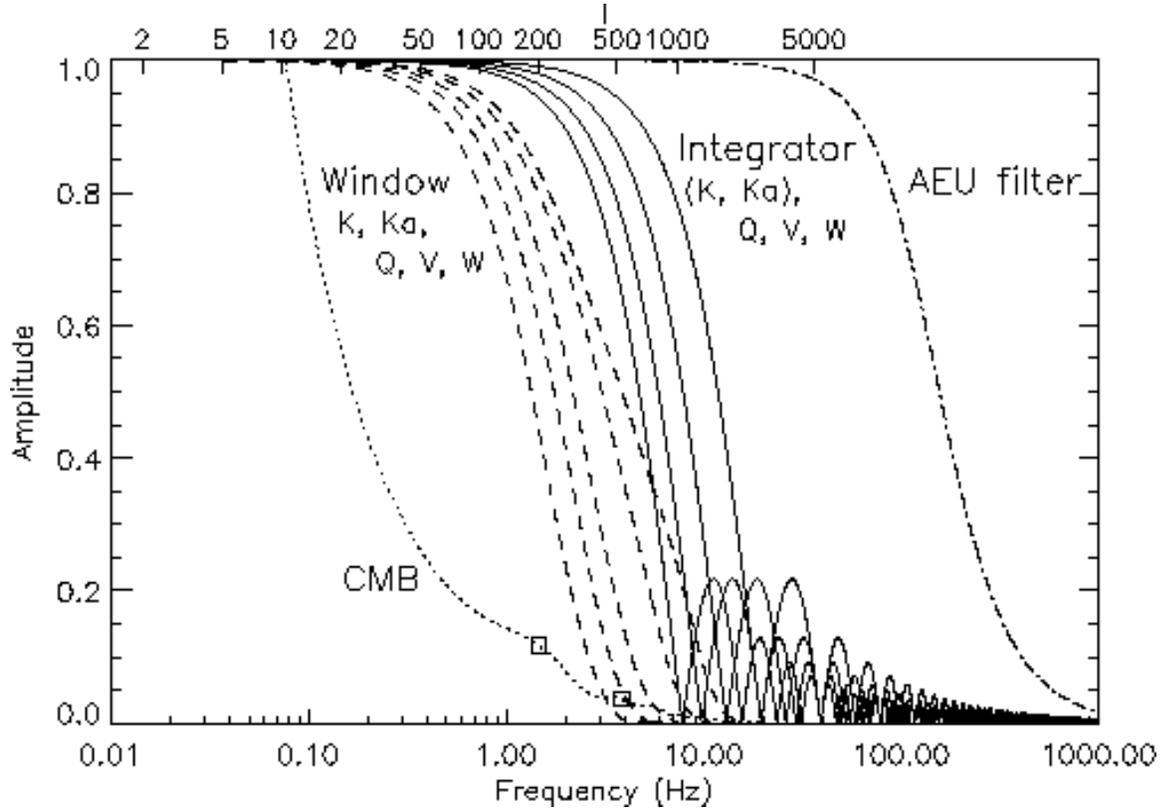}

\caption{ Frequency response of the \wmap~data collection system and some representative
input spectra. The curve labeled ``AEU filter'' (dot-dash)
is the amplitude response of the 2-pole Bessel filter ($f_{\rm 3dB} = 100$~\Hz) before the voltage to
frequency converter in the Analog Electronics Unit. The 4 solid lines labeled ``Integrator'' are the absolute values
of the \sinc~ functions
corresponding to the $128, 102.4, 76.8$ and $51.2$ \ms integration periods used for the different
radiometer frequency bands. (See Figure 8 of \citet{bennett/etal:2003}.) The dashed lines labeled 
``Window'' are the frequency spectra
that would result from \wmap's scan pattern and beams if it were observing a scale invariant angular power spectrum
($C_l = const.$) sky based on the window functions in \citet{page/etal:2003b}. The dotted line labeled CMB is the input signal to the data collection system
which would occur for a typical CMB sky sampled by an ideal pencil beam. The two squares on this
last curve show the approximate locations of the first two peaks in the CMB power spectrum. The
response of the AEU filter is down from unity by $0.1$\% at $l = 1000$, indicating that it has a negligible
effect on the shape of the observed power spectra.  }
\label{fig:data_col}
\end{figure*}

The key radiometer performance parameters are the frequency bandpasses, data collection system
responses, and radiometer noise and gain characteristics. The frequency bandpasses and
data collection
system  were fully characterized in pre-launch ground tests and are
 described in \citet{jarosik/etal:2003}. Figure~\ref{fig:data_col} summarizes the
audio frequency response of the data collection system together with the frequency roll-offs 
associated with the beam sizes and the angular spectrum of a typical CMB signal.
The radiometer gain and noise properties were also characterized during ground tests, but the final
precise measurement of these quantities is done while on orbit since they depend on details of 
the thermal environment of the radiometers which could not be predicted  precisely before launch.

\section{THE RADIOMETER THERMAL ENVIRONMENT}
Approximately 2 hours after launch, power was applied to the \wmap~radiometers while the
Focal Plane Assembly (FPA) components were at approximately 265 K; all 20 radiometers
were observed to be operational. (The FPA comprises the input stages of the radiometers
and feed horns and is designed to passively cool to $\approx 90$ K.) Over the next 2 days all the
major radiometer assemblies, the FPA,
 the Receiver Box (RXB),
the Analog Electronics Unit (AEU) and the Power Distribution Unit (PDU)
quickly approached nominal operating temperatures and radiometer performance
improved as expected. (See Figure 7 of \citet{bennett/etal:2003} for an block diagram
containing details of these units.)  Infrared irradiation from the Earth and
Moon during the first month of the mission, the trajectory phasing loops, induced thermal fluctuations in
the instrument components, limiting the utility of radiometer performance
analyses which could be performed. During this period, however,   valuable measurements of beam sidelobe pattern were
 obtained.
These large  thermal perturbations ended after the lunar encounter on 2001 July 30, allowing the temperatures 
of the instrument components to approach their final steady state values. All
performance data presented here were taken after 2001 August 10, 
by which time the satellite temperatures had stabilized and data for
the first year sky maps were being taken. 

\wmap~is outfitted with 57 high resolution platinum resistance thermometers that monitor the
temperature of representative instrument components. Each thermometer is read out every
23 seconds, has  $0.5~\mkelvin$
resolution, $\pm 1$~K accuracy, and $\approx 1$ least significant bit of readout noise to allow averaging of multiple
samples.
There are 17 such thermometers on FPA components, 12 on RXB components, 13 on
the optics\footnote{One of the thermometers attached to the top of a radiator panel
is non-functional. See~\citet{limon/etal:2003} for details.}, 9 in the AEU and 6 in the PDU.

Table~\ref{table:thermal_perf} lists the 12-month mean temperatures,  
peak-to-peak temperature variations, 
and limits on the measured spin synchronous temperature variations of the AEU, FPA, PDU and RXB.
 The total variation also contains a slow annual term arising
from the ellipticity of the Earth's orbit modulating the level of insolation on the spacecraft plus
 a contribution
from the initial thermal transient associated with phasing loops and the final lunar encounter. Details
of the thermal history of each component are presented in \citet{limon/etal:2003}.
Future total variations are predicted to be $\approx 1/3 - 1/2$ the values given in the Table since the
transient effects will not be repeated.
  The spin synchronous
term  was measured by removing  a slowly varying baseline from the temperature data 
and binning the residual temperature fluctuations synchronously with the
spacecraft spin relative to the Sun direction. Keeping the spin synchronous temperature fluctuations at or below
the design requirement values is essential in order to meet the systematic error
requirements~\citep{jarosik/etal:2003}.
The measured spin synchronous temperature variations of all the instrument components are at least a factor
of \emph{50 smaller} than the maximum allowed values specified in the systematic error budget,
the values of which are also presented in  Table~\ref{table:thermal_perf}.

\begin{deluxetable}{ccccc} 
\tablecaption{\wmap~Radiometer Thermal Environment Summary\label{table:thermal_perf} }
\tablecomments{ This Table summarizes the thermal environment of the \wmap~
radiometers. The ranges specified for the ``Mean Values'' are  12 month averages of the
coldest and hottest thermometer in each assembly. The ``Total Variation'' values are the
largest variations  observed for any thermometer within the corresponding assembly.
The measured spin synchronous values are obtained by
binning measured temperatures  in  spin synchronous coordinates relative
to the Sun, after spline removal of long term temperature drifts. The values presented are
derived from one month of data and are the largest limits obtained for any thermometer 
within each assembly. No significant spin synchronous temperature variations are observed,
so the limits presented are determined by the binned readout noise of the thermometers. The measured
limits on the spin synchronous temperature variations are at least a factor of 50 smaller
than the maximum values allowed by the pre-flight systematic error budget. }
\tablecolumns{4}
\tablehead{
\colhead{Assembly}&
\colhead{Mean Values}&
\colhead{Total Variation}&
\colhead{Requirement spin-sync}  &
\colhead{Measured spin-sync}  \\ 
& (K) & (K p-p) & (\ukelvinrms) &  (\ukelvinrms)  \\

} 
\startdata
PDU    &$295.9 - 298.1$ & 2.64  &$ < 10000$ & $< 7.5$ \\ 
AEU    &$298.7-305.9$   & 2.52  &$ < 10000$ & $< 7.8$ \\
RXB    &$286.4-288.5$   & 1.20  &$ < 500$  & $< 7.6$ \\
FPA    &$87.9 - 91.0$   & 0.88  &$ < 500$  & $< 6.4$ \\
\enddata

\end{deluxetable}

\section{RADIOMETER PERFORMANCE}
\subsection{Sensitivity}
The radiometers must be calibrated using known input signals  
to determine their sensitivities. 
During ground testing at the Goddard Space Flight Center (GSFC), calibrations were performed with 
thermally regulated
full aperture loads attached to the feed horn inputs.
On-orbit, the $3~\mkelvin$ CMB dipole signal, coupled into the radiometer through the full
optical system, provides a stable  calibration signal and is used to
track calibration drifts. Absolute calibration is provided by the
annual modulation of the dipole signal arising from the known orbital
velocity of the Earth about the Sun, and the CMB monopole temperature
as determined by \citet{mather/etal:1999}. The conversion factors presented 
in \citet{jarosik/etal:2003} are used to convert the dipole calibrations into  Rayleigh-Jeans units.

Table \ref{table:final_perf} contains the radiometer sensitivities 
measured at GSFC and on-orbit.  The GSFC sensitivities  have been scaled to compensate
for the difference between the radiometer environments during  the ground and on-orbit measurements. 
The sensitivity values measured
during the GSFC test agree fairly well with the values measured on-orbit. 
The K (23 \GHz), Ka (33 \GHz), Q (41 \GHz), and V (61 \GHz) band
radiometers seem to have slightly lower noise on-orbit than predicted from the scaled GSFC
results. This is thought to be the result of small reflections in the feed horn loads used during
the ground test, the effects of which were not included in the analysis. 
The W (94 \GHz) band radiometers appear slightly
noisier than predicted. In this case it is thought that the simple
power law scaling used to predict the cryogenic HEMT noise temperature
 based on its physical temperature is inaccurate.
Table~\ref{table:final_perf} also contains the estimated  mean pixel noise values 
($3.2\times 10^{-5}$ sr/pixel), derived from the measured on-orbit noise levels,
for maps created by combining all the radiometer channels in each frequency band
for the originally approved 2 year mission and the currently approved extension to a 4  year mission.

\begin{deluxetable}{ccccccccc} 
\tablecaption{Radiometer Performance Summary\label{table:final_perf} }
\tablecomments{ Summary of the sensitivities, $f_{\rm knee}$, $T_{\rm off}$,
of the 20 radiometers comprising \wmap~as measured during
integration and testing (GSFC) and on-orbit at Earth-Sun L2.
 The sensitivity values given are for the 
combined output of the two detectors
on each radiometer, e.g. K113 + K114. The values of $\Delta T/{\rm pixel}$ are for all radiometers
in each frequency band combined. The sensitivity values from
 ground tests have been scaled to approximate those 
expected on-orbit for an FPA temperature of 89~K. The pixel noise estimate assumes
4 (2) years of data, $3.2 \times 10^{-5}$ sr/pixel, and uniform sky coverage. The last column contains
limits on spin synchronous artifacts in the time ordered data, $\Delta T_{\rm ss}$,  obtained from combining
the products of the temperature susceptibility coefficients measured during grounds tests with the
spin synchronous temperature fluctuation limits from Table~\ref{table:thermal_perf}.  These data 
indicate that the radiometers are functioning properly and that spin synchronous artifacts in the
time ordered data are expected to be much smaller that the maximum values permitted ($2.8~\ukelvin$)
in the systematic error budget. All temperatures are in Rayleigh-Jeans units. }
\tablecolumns{8}
\tablehead{
\colhead{Radiometer}&
\multicolumn{2}{c}{$\rm{Sensitivity}$} &
\colhead{$\Delta T/{\rm pixel}$} &
\multicolumn{2}{c}{ $f_{\rm knee}$ }  &
\multicolumn{2}{c}{ $T_{\rm off}$ }  &
\colhead{$\Delta T_{\rm ss}$}  \\ 
&
\multicolumn{2}{c}{$(\mkelvin~\sec^{1/2})$} &
(\ukelvin) &
\multicolumn{2}{c}{(\mHz)}&
\multicolumn{2}{c}{(K)}&
(\ukelvinrms)
\\
 & GSFC & Flight & 4 yr (2 yr) &GSFC & Flight &GSFC & Flight & Flight
} 
\startdata
K11     & 0.72 & 0.66& 19.4  (27.5)  & 6.13&  $ 0.40 $ & -0.033  & $0.19$ & $<0.02$ \\ 
K12     & 0.87 & 0.75&      & 5.37&  $ 0.51 $ & -0.249  & $0.22$ & $<0.02$\\
Ka11    & 0.75 & 0.71& 19.9  (28.2)  & 1.66&  $ 0.71 $ &  0.387  & $0.49$ & $<0.03$ \\
Ka12    & 0.77 & 0.72&      & 1.29&  $ 0.32 $ &  0.072  & $0.25$ & $<0.02$\\ 
Q11     & 0.99 & 0.92& 20.5  (29.0)  & 3.21&  $ 1.09 $ &  0.092  & $-0.11$ & $<0.04$\\
Q12     & 0.95 & 1.02&      & 3.13&  $ 0.35 $ &  0.142  & $0.12$ & $<0.02$\\
Q21     & 0.89 & 0.85&      & 1.92&  $ 5.76 $ &  0.552  & $1.14$ & $<0.03$\\ 
Q22     & 1.04 & 0.99&      & 4.61&  $ 8.62 $ &  1.036  & $1.54$ & $<0.15$\\ 
V11     & 1.25 & 1.22& 24.0  (34.0)  & 2.56&  $ 0.09 $ & -0.448  & $0.06$ & $<0.01$\\
V12     & 1.07 & 1.11&      & 4.49&  $ 1.41 $ & -0.270  & $0.24$ & $<0.04$ \\
V21     & 1.01 & 0.97&      & 2.43&  $ 0.88 $ & -0.265  & $0.31$ & $<0.04$\\
V22     & 1.13 & 1.10&      & 3.06&  $ 8.35 $ &  0.352  & $0.95$ & $<0.01$\\ 
W11     & 1.18 & 1.35& 23.1  (32.7)  & 16.17& $ 7.88 $ & -0.451  & $1.40$ & $<0.01$ \\
W12     & 1.41 & 1.61&      & 15.05& $ 0.66 $ & -2.064  & $-0.10$ & $<0.09$ \\
W21     & 1.38 & 1.61&      & 1.76&  $ 9.02 $ & -0.091  & $1.41$ & $<0.02$\\
W22     & 1.44 & 1.72&      & 0.77&  $ 7.47 $ &  0.008  & $1.53$ & $<0.01$\\
W31     & 1.47 & 1.65&      & 1.84&  $ 0.93 $ & -1.151  & $-0.39$ & $<0.04$\\
W32     & 1.69 & 1.86&      & 2.39&  $ 0.28 $ & -1.117  & $0.05$ & $<0.17$\\
W41     & 1.60 & 1.71&      & 8.46&  $ 46.5 $ &  1.300  & $3.33$ & $<0.02$\\
W42     & 1.43 & 1.65&      & 5.31&  $ 26.0 $ &  1.441  & $3.27$ & $<0.12$\\
\enddata

\end{deluxetable}
\begin{figure*}
\includegraphics[width=6in]{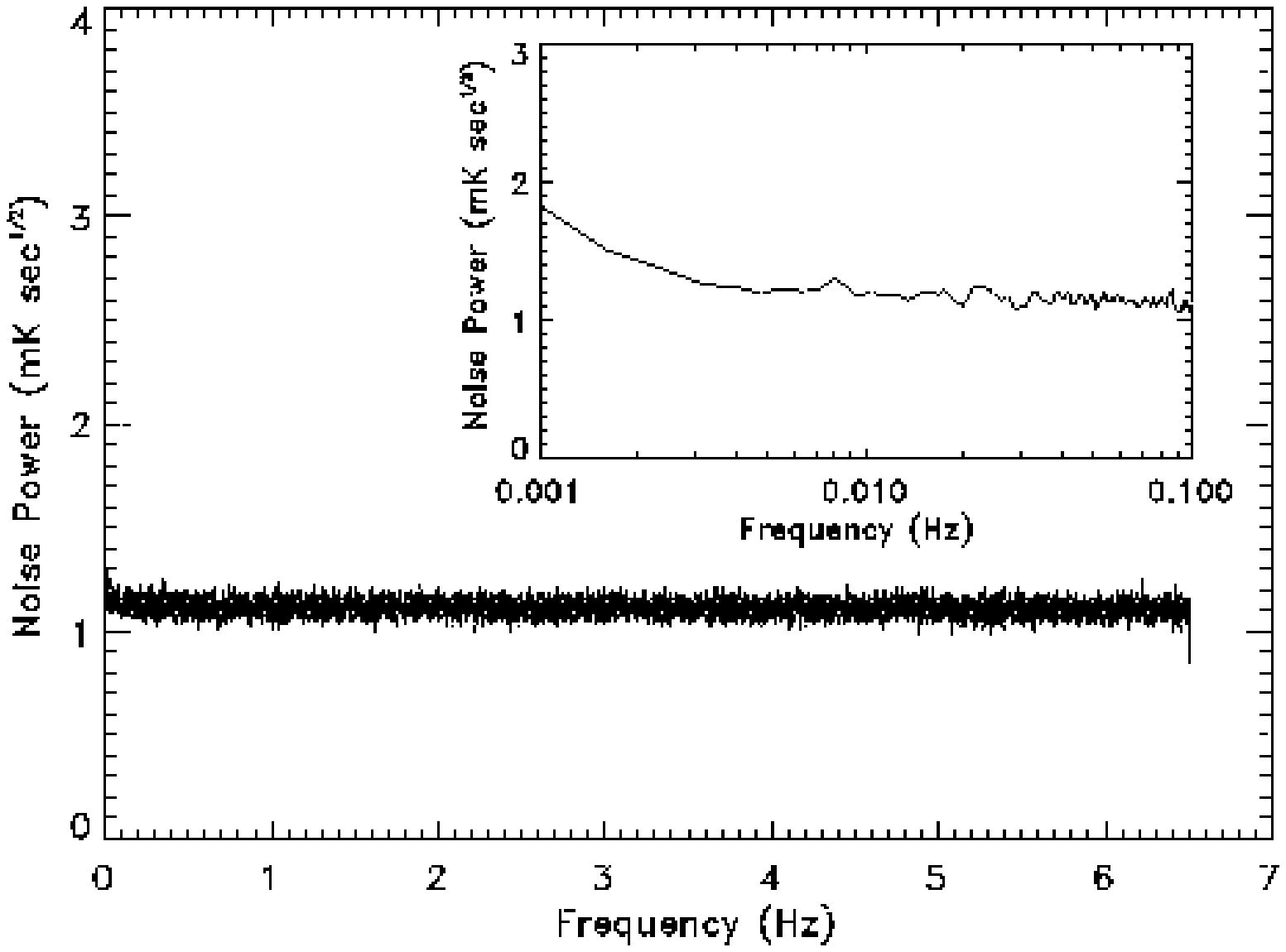}
\caption{ Noise power spectral density of the V12 radiometer obtained from 3 days of on-orbit data.
 Sky signals arising from the dipole, CMB, Galaxy and point sources have been removed. The inset 
contains an expanded
view of the low frequency region of the same data. The flat spectrum down to very low
frequencies ($f_{\rm knee} = 1.41~\mHz$) indicates
 proper radiometer performance.
}\label{fig:noise_spec}
\end{figure*}

\subsection{ Stability }

Ideally the radiometer noise has a flat (white) power spectrum over the
audio bandwidth of the signal being measured.
Given the \wmap~beam sizes and scan rates, the signal bandwidth extends from $\approx 0.008 - 8~\Hz$.
 Figure~\ref{fig:noise_spec}
shows a power spectrum of the V12 radiometer noise obtained on-orbit.  (See \citet{jarosik/etal:2003} for
an explanation of the radiometer labeling conventions.)
The noise power spectrum is very flat and only turns up
at very low frequencies. The increase in noise at low frequency is characterized by
a ``1/f knee frequency'', $f_{\rm knee}$, taken to be the value at which
the noise power spectral density  increases to $\sqrt{2}$ times its high frequency value, in this case
$f_{\rm knee} = 1.12$ \mHz.
Table~\ref{table:final_perf} contains the values of $f_{\rm knee}$ measure on-orbit for all the radiometers.
Most of the radiometers have  $f_{\rm knee}$  values below $8~\mHz$,
the low frequency end of the
signal bandwidth, but a few are problematic, particularly W41 and W42. 

\begin{figure*}
\epsscale{1.0}
\includegraphics[width=6in]{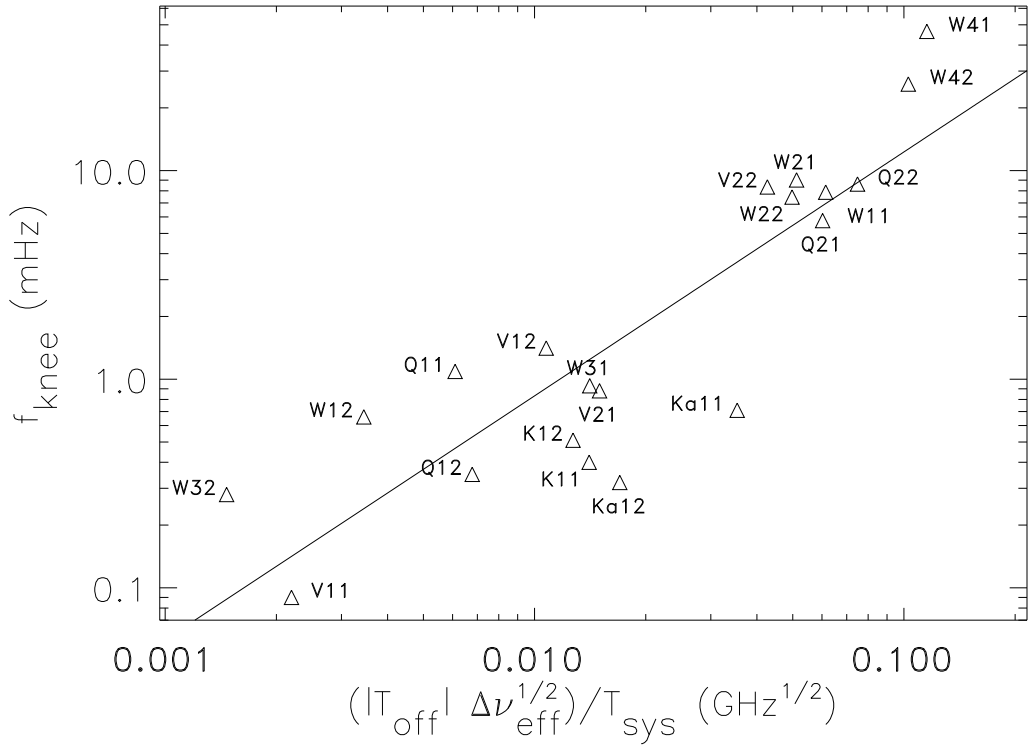}

\caption{ Dependence of $f_{\rm knee}$ on $T_{\rm off}$ for the 20 radiometers
comprising \wmap.  The solid line is
a power law fit to the data of the form $f_{\rm knee} \propto ((|T_{\rm off}|\Delta\nu^{1/2}_{\rm eff})/T_{\rm sys})^{2/\alpha}$
 with $\alpha = 1.70$. The scaling of $f_{\rm knee}$ with  $T_{\rm off}$ indicates that $f_{\rm knee}$ is largely
determined by radiometer gain fluctuations modulating
the signal from the radiometer offsets,  as expected.   }
\label{fig:f_knee}
\end{figure*}

The radiometer offset temperatures,  $T_{\rm off}$,  measured at GSFC and on-orbit  are also
presented in Table~\ref{table:final_perf}.
These values correspond to the size of the radiometer output signal
when both inputs are observing regions with the same antenna temperatures.
(See \citet{jarosik/etal:2003} for the precise definition of $T_{\rm off}$ used for \wmap.) 
A significant
correlation between the magnitude of the radiometer offset, $T_{\rm off}$, and value of $f_{\rm knee}$ is
 expected for differential radiometers; this correlation 
arises from radiometer gain fluctuations modulating the radiometer
offset signal. If it is assumed that all the amplification chains in the radiometers
have power gain fluctuations of the same fractional amplitude and frequency spectra proportional
to $f^{-\alpha}$, then the $f_{\rm knee}$ is expected to
scale as $f_{\rm knee} \propto (|T_{\rm off}|\Delta\nu^{1/2}/T_{\rm sys})^{2/\alpha}$. Here $T_{\rm sys}$ and  $\Delta \nu$
are the radiometer system noise temperatures and noise equivalent bandwidths. Figure~\ref{fig:f_knee}
shows on-orbit data 
for the 20 \wmap~radiometers. The value of the exponent $\alpha$ measured from these data is $1.70$,
 somewhat larger that the value of $0.96$ reported by \citet{wollack/pospieszalski:1998}. Direct
comparison of these values of $\alpha$ is complicated by the fact that the data contained in
Figure~\ref{fig:f_knee} comes from 20 different radiometers fabricated from HEMT devices with
three different geometries, while the \citet{wollack/pospieszalski:1998} results were obtained 
from measurements on a single amplification chain.  The difference between the measured values of $\alpha$
may therefore not be significant. 

The source of the high value of $f_{\rm knee}$ in the W41 and W42 radiometers is 
the large  $|T_{\rm off}|$ for these radiometers. The on-orbit values of $|T_{\rm off}|$
are significantly larger than the values measured during the GSFC tests for the W21, W22, W41
and W42 radiometers.
 The fact that the radiometer $f_{\rm knee}$
values track the measured values of $|T_{\rm off}|$, as shown in Figure~\ref{fig:f_knee}, indicates
the radiometers are operating as expected. The exact source of the larger $|T_{\rm off}|$ values is
not known.

Even with the relatively low values of $f_{\rm knee}$ for the majority of the radiometers,
noise correlations are not negligible.
Maps produced directly from radiometer data for radiometers with $f_{\rm knee} \lesssim f_{\rm spin} = 7.75~\mHz$
exhibited small but detectable  pixel noise correlations (``stripes'') which could corrupt
power spectrum determinations.  
In order to
minimize these effects in the final sky maps, a pre-whitening procedure
has been adopted which fits a baseline to the TOD after removal
of an estimated sky signal obtained from the sky maps. Details of the pre-whitening
process are presented in \citet{hinshaw/etal:2003b}.

The last column in  Table~\ref{table:final_perf} presents limits to spin synchronous artifacts in the
TOD for each of the radiometers. These values were determined by multiplying the 
limits on the spin synchronous
temperature fluctuations of each radiometer section (AEU, PDU, FPA and RXB) by
 the appropriate susceptibility
coefficient determined during ground testing, and summing values in quadrature. 
\citet{hinshaw/etal:2003b} obtain similar results using susceptibility coefficients
derived from on-orbit data.
In all cases the 
limits on spin synchronous artifacts are much smaller than the $2.8~\ukelvin$ allocated 
to the radiometers
and data collection system in the systematic error budget \citep{jarosik/etal:2003}. 

  It is possible to obtain an estimate of spin synchronous
artifacts directly from the TOD independent of any model of its source. This was
accomplished by binning 40 days of TOD (after removal of the CMB dipole signal) 
in an azimuthal coordinate about the spin axis of the
spacecraft.   
(Recall the Sun's elevation is fixed during observations by the \wmap~scan pattern.)
Each TOD point was binned according  to the Sun's location at the time of the
observation, and bin averages calculated.
TOD from  all of the V and W band radiometers was used and
 a $10^\circ$ Galactic cut applied. The 
limit obtained by this technique is $\Delta T_{\rm ss} < 0.14~\ukelvin $ rms, 
well below the $9~\ukelvin$ value allowed in the systematic error budget~\citep{jarosik/etal:2003}.
Note that this limit includes any artifacts which originate in the radiometers, data collection
system  or optical
system. Since both the estimated and  measured systematic error
limits are so small,
 \emph{no corrections arising from spin synchronous radiometric artifacts 
are applied to the TOD in the production of the one year sky maps.}

During the data period incorporated in the 1 year maps a total of 21  sudden jumps
(``glitches'') have occurred in the outputs of the radiometers.  Jumps were exhibited by the
 Ka11, Ka12, Q11, Q12 and W12
radiometers. Of these events, 19 involved  single radiometers, one simultaneously
affected two radiometers and one affected three.  Similar  
artifacts were  observed during ground testing and have been identified  as  small
parameter shifts in the properties of several microwave components resulting from sudden
releases of internal mechanical  stresses. The most common events involve the phase 
switches and filters, which are constructed with suspended stripline elements, but events have
also been observed involving waveguide components connecting the orthomode transducers
to the inputs of the radiometers. These events
result in a negligible change in radiometer performance, but must be  identified and 
excised from the TOD  in production of the maps. Although the duration of each event
is short, $< 1$~s, the data processing pipeline~\citep{hinshaw/etal:2003b} requires removal
of several hours of data surrounding each event resulting in a data loss of $\approx0.13 \%$.
 A complete tabulation 
of these events can be found in  \citet{limon/etal:2003}.

\subsection{ Input Transmission Imbalance}
Ideally the output of each  \wmap~radiometer would exhibit a purely differential
response to sky signals, rejecting any common mode component. 
This is not true if the power  transmission coefficients, $\alpha_{\rm A}$ and $\alpha_{\rm B}$,
of the optics and waveguide components, which couple the sky signals from the A and B
side beams to the radiometer, are not equal.
In this case the output of a radiometer, $S$,  when the beams observe regions of
sky with antenna temperatures  $T_A$ and $T_B$  will be
\begin{equation}
S = G(\alpha_{\rm A} T_A - \alpha_{\rm B} T_B),
\end{equation}
where $G$ is the gain of the radiometer.

\begin{figure*}
\epsscale{1.0}
\includegraphics[width=6in]{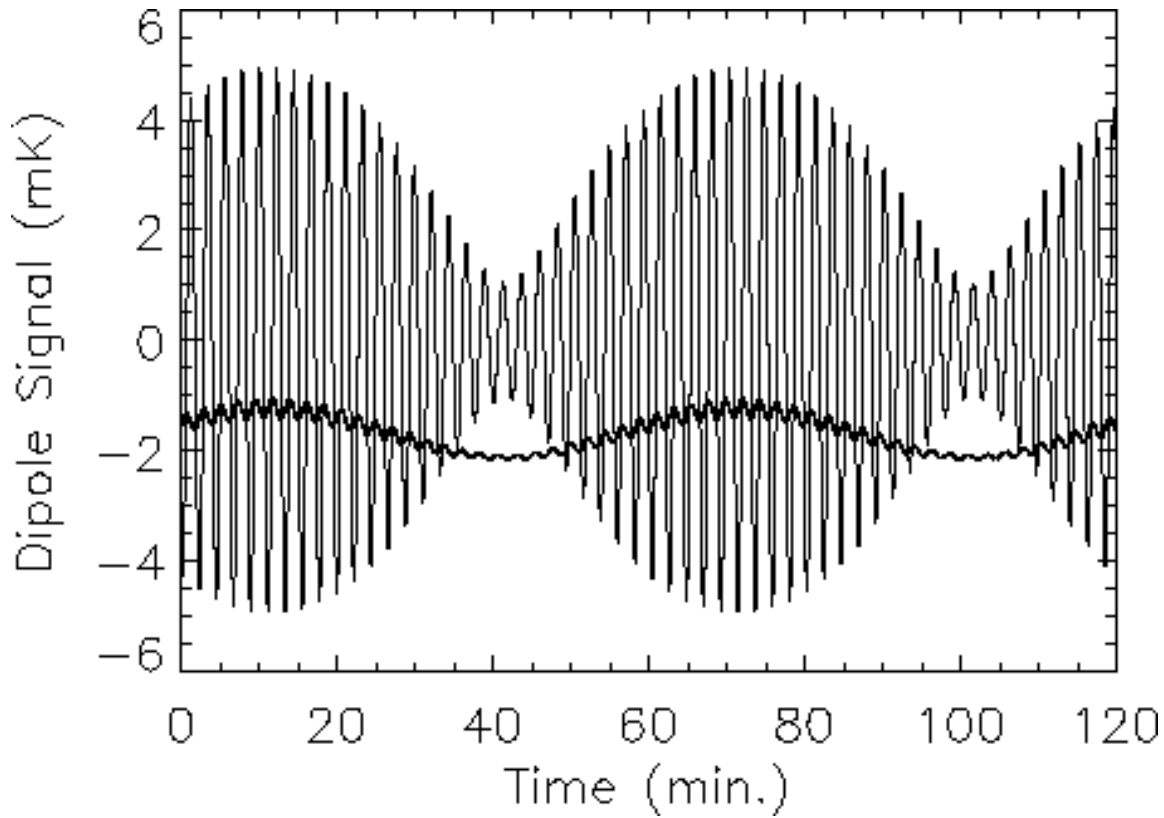}

\caption{ Predicted differential (thin) and common mode (thick) CMB dipole signal for a typical 2 hour
period of \wmap~observations. The rapid period ($\approx 2$ minute) corresponds to \wmap's
spin rate and the slower period ($1$ hr) to the precession rate. Measurement of the common mode signal
component in a radiometer's output allows characterization of the  non-ideal (common mode) radiometric 
response.}
\label{fig:dipole_template}
\end{figure*}

Using the CMB dipole signal it is possible to measure the common
mode response of a radiometer to sky signals and therefore infer the value of $\alpha_{\rm A} - \alpha_{\rm B}$.
Figure~\ref{fig:dipole_template} shows the pure predicted differential and common mode
CMB dipole signals, $T_d = T_A - T_B$ and $T_c = T_A + T_B$, for a typical 2 hour period of \wmap~observations. 
Note that $T_d$ and $T_c$ are very nearly orthogonal when averaged over a precession period due
to the symmetry of \wmap's scan pattern. \citep{bennett/etal:2003}
 The common
mode signal is comprised of three terms. The first appears as the $-1.8~\mkelvin$ offset in the figure
and is modulated on the one year timescale as \wmap's precession axis changes orientation with
respect to the CMB dipole axis. This signal component is at such a low frequency and amplitude
that it cannot be distinguished from long term drifts in the radiometer offsets. The other two
terms vary on the timescale of the precession (1 h) and the spin (129 s), with
most of the power being in the lower frequency component. The amplitude of these two signals combined
is $\approx 0.5~\mkelvin$, so a 1\% common mode response corresponds to
 a signal amplitude of $\approx 5~\ukelvin$. 

The common mode radiometric response was measured by fitting the raw radiometer TOD to
a function of the form
\begin{equation}
S(t_j) = \sum_{\rm i=0}^{3} c_i (t_j-t_0)^i + \beta_c T_c(t_j) + \beta_d T_d(t_j) \label{eqn:fit_func},
\end{equation}
where $t_j$ is the time of the observations, $c_i$ are polynomial coefficients
used to remove a baseline, $T_c$ and $T_d$ are the common mode and differential
CMB dipole signal templates and $\beta_C$ and $\beta_S$ are the coefficients to be fit. Raw data
must be used since the calibration and baseline fitting procedures \citep{hinshaw/etal:2003b} remove
most of the very low frequency  ($\approx 0.3~\mHz$) signal we are attempting to measure.
The measured values of $\beta_c$ and $\beta_d$ are related to the transmission
coefficients $\alpha_{\rm A}$ and $\alpha_{\rm B}$ by
\begin{eqnarray}
\beta_c = \frac{G}{2}(\alpha_{\rm A} - \alpha_{\rm B}) \\
\beta_d = \frac{G}{2}(\alpha_{\rm A} + \alpha_{\rm B}).
\end{eqnarray}
 The TOD is divided into segments containing
an integral number of precession periods, $n_{\rm prec}$, and a function of the form given by 
 Equation~\ref{eqn:fit_func} fit to each segment. Observations at Galactic latitude $|b| < b_{\rm min}$ are
excluded from the fits. Values of the fractional transmission imbalance, $x_{\rm im}$, defined as
\begin{equation}
x_{\rm im} \equiv \frac{\langle\beta_c\rangle}{\langle\beta_d\rangle} = \frac{\langle\alpha_{\rm A} -
 \alpha_{\rm B}\rangle}{\langle\alpha_{\rm A} + \alpha_{\rm B}\rangle}
\approx\frac{\alpha_{\rm A} - \alpha_{\rm B}}{2},
\end{equation}
are presented in  Table~\ref{table:diff_trans}.
 The braces indicate a  weighted average of the measured coefficients, $\beta_c$ and $\beta_d$, 
over the 232 day period used in the analysis.
The results of the fits are insensitive to the values of $n_{\rm prec}$ chosen in the range of 4-10, indicating
that the baseline removal polynomial is not removing a significant amount of common mode signal.
The fit values are also insensitive to the value of $b_{\rm min}$ used provided $b_{\rm min}> 10^\circ$, 
indicating that Galactic signals are not biasing the results.

\begin{deluxetable}{lllll}
\tablewidth{4in}
\tablecaption{ Input Transmission Imbalance Measurements of the \wmap~Radiometers\label{table:diff_trans}}
\tablehead{
\colhead {Radiometer} & \colhead{$x_{\rm im}$} & \colhead{\phantom{xxxxx}} &
\colhead{Radiometer} &\colhead{$x_{\rm im}$}} 
\startdata
K11&  -0.00204&  &K12&-0.00542 \\
Ka11&  0.00115&  &Ka12& 0.00108 \\
Q11&  -0.00200&  &Q12& 0.00010 \\
Q21&   0.01251&  &Q22& 0.01433 \\
V11&  -0.00354&  &V12&-0.00015\\
V21&   0.00682&  &V22& 0.00598 \\
W11&   0.00846&  &W12& 0.00581\\
W21&   0.01550&  &W22& 0.00849 \\
W31&   0.00253&  &W32& 0.00542 \\
W41&   0.01536&  &W42& 0.01581 
\enddata
\tablecomments{ 
Measurement of the fractional input transmission imbalance,  $x_{\rm im}$,
obtained from 232 days of data with $n_{\rm prec} = 10$ and $b_{\rm min} = 15^\circ$. They were obtained
via measurements of the radiometer responses to the common mode signal arising from the CMB dipole.
All the values are small, nevertheless corrections for this effect have been included in the
map making algorithm.}
\end{deluxetable}

\begin{figure*}
\epsscale{1.0}
\includegraphics[width=6in]{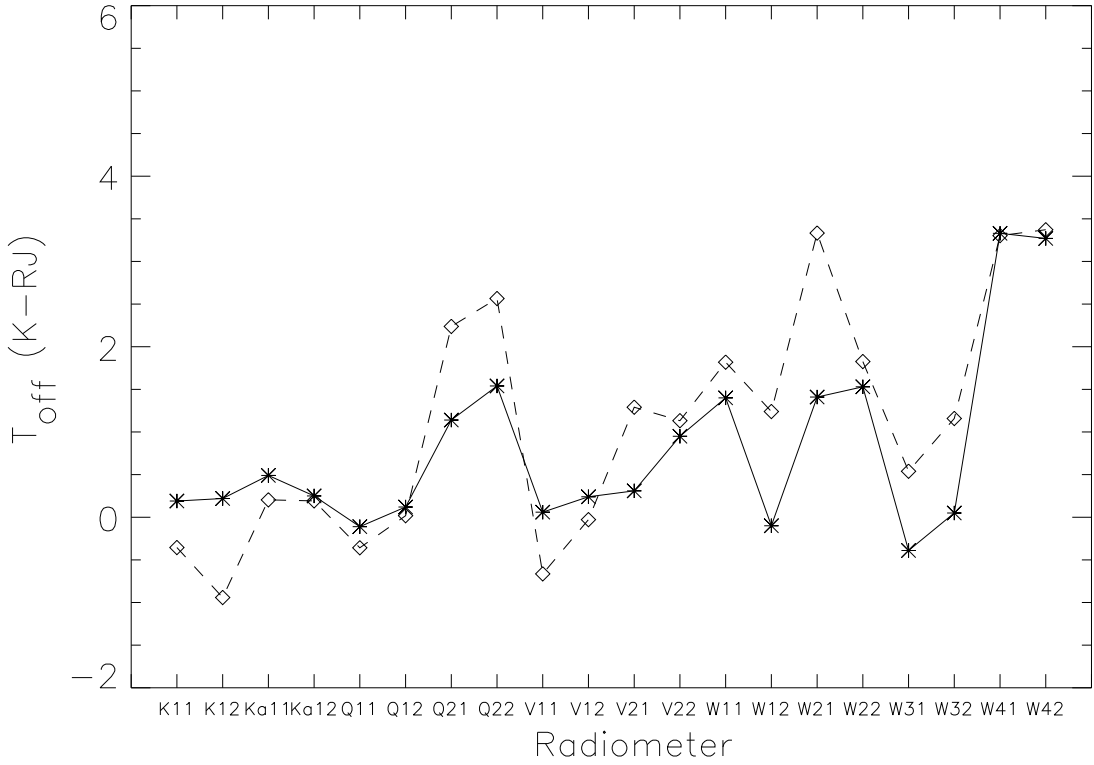}

\caption{ On-orbit measurements of the radiometer offset temperatures, $T_{\rm off}$ (stars), 
and predicted contributions to
the offset temperature, $\Delta T_{\rm off}$ (diamonds), for all 20 \wmap~radiometers. Values
of $\Delta T_{\rm off}$ are obtained from measurements of the radiometer common mode responses to the 
CMB dipole signal. Significant correlation between the measured offset and the predicted
values of $\Delta T_{\rm off}$ are expected and observed. The lines are drawn to aid in seeing the
correlation.}
\label{fig:toff_and_dloss}
\end{figure*}

A differential loss in the input waveguides and optics should also contribute to radiometric
offsets, since the lossy components will also emit radiation. The contribution to the radiometer
offsets arising from such a differential loss can be estimated as
\begin{eqnarray}
\Delta T_{\rm off} & = & [(\alpha_{\rm A} T_{\rm CMB} + ( 1 - \alpha_{\rm A})T_{\rm FPA}) -  (\alpha_{\rm B} T_{\rm CMB} + ( 1 - \alpha_{\rm B})T_{\rm FPA})\\
	& = &(\alpha_{\rm A} - \alpha_{\rm B})(T_{\rm CMB}-T_{\rm FPA}) \\
	&\simeq& 2 x_{\rm im}(T_{\rm CMB}-T_{\rm FPA}),\label{eqn:toff_pred}
\end{eqnarray}
where it has been assumed that all the loss occurs at physical temperature $T_{\rm FPA}$ and
$\alpha_{\rm A} \simeq \alpha_{\rm B} \simeq 1$. This effect is expected to have a significant contribution
to the total $T_{\rm off}$ of each radiometer. Figure~\ref{fig:toff_and_dloss} displays the values
of $T_{\rm off}$ measured in flight and those predicted from Equation~\ref{eqn:toff_pred}, assuming
$T_{\rm FPA} - T_{\rm CMB} = 86$~K. A significant correlation is evident between the values predicted 
by this model and the observed values, supporting the validity of the technique. 
	
The model of the radiometer response to sky signals used in the map making procedure
\citep{hinshaw/etal:2003b} is
\begin{eqnarray}
S& =& G\left[ \frac{\alpha_{\rm A} - \alpha_{\rm B}}{2} T_{\rm c} + \frac{\alpha_{\rm A} + \alpha_{\rm B}}{2} T_{\rm d}\right] \\
S& =& G'[(1 + x_{\rm im}) T_{\rm A} - (1-x_{\rm im}) T_{\rm B}],
\end{eqnarray}
where the mean of the input transmission coefficients, $(\alpha_{\rm A} + \alpha_{\rm B})/2$,
 has been
absorbed in the measured  gain, $G'$, determined by fitting the differential CMB dipole
signal.

\subsection{ Noise Statistics }
\begin{figure*}
\includegraphics[width=6in]{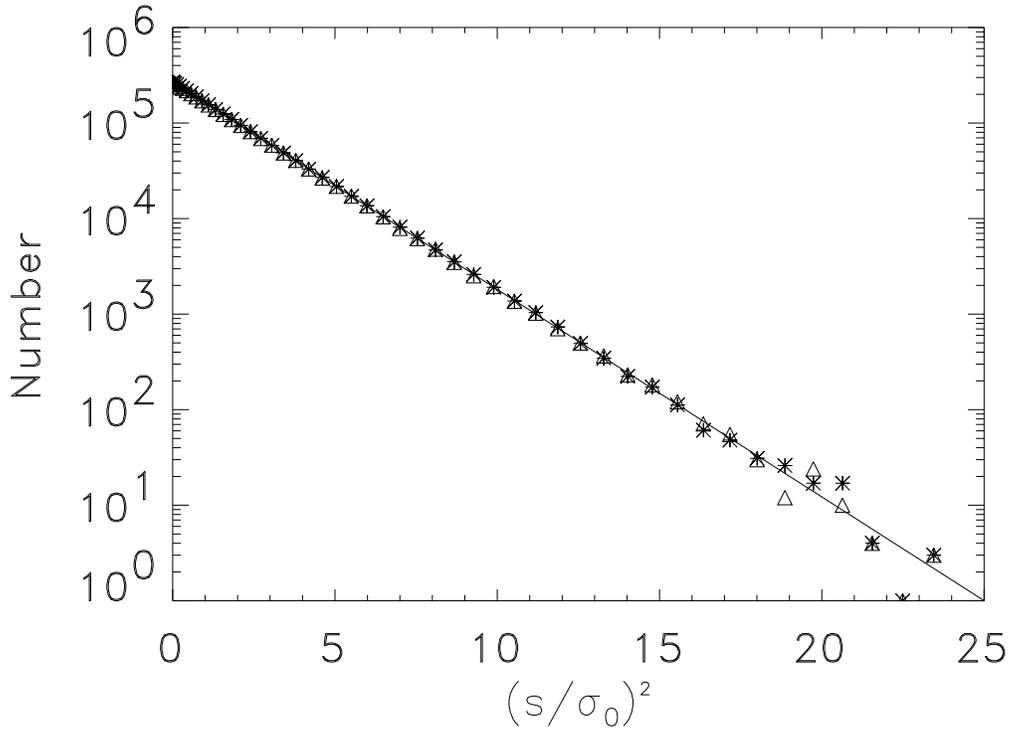}

\caption{ Distribution  of the pre-whitened V12 radiometer noise obtained
from ten days of observations. Sky signals arising from the dipole, CMB and
Galaxy have been removed. Data points were cut when either radiometer
beam encountered a planet or a region of high Galactic emission. The line
corresponds to a unit variance Gaussian distribution normalized to the observed
frequency at $s = 0.$ The two symbols denote the values obtained from the
two sides of the distribution. The highly Gaussian distribution indicates that
 the
noise variance for each pixel of the resulting sky maps should scale inversely
with the number of observations of that pixel. }
\label{fig:noise_dist}
\end{figure*}

The intrinsic radiometer noise is expected to exhibit a Gaussian distribution.
 A plot of the  V12 radiometer noise distribution  
obtained from 10 days of data  is presented in Figure~\ref{fig:noise_dist}. This data period
was chosen to be free of disturbances (such as orbital station keeping maneuvers) and its duration
short enough so that the slow variations in system noise, driven by the annual
temperature variation of the FPA, could be ignored.  The noise signal was
obtained by removing the CMB dipole and an estimated sky signal (obtained from the final maps)
from the pre-whitened TOD. Data when either of the beams was in a region
of high Galactic emission (as determined by the Kp4 mask \citep{bennett/etal:2003b}) or near a planet were cut.
No other cuts were made. The distributions for all 20 
radiometers exhibited similar behavior, following  Gaussian distributions over at least
5 decades. No statistically significant skewness or kurtosis is observed in the
noise distributions of any of the radiometers.

\begin{figure*}
\includegraphics[width=6in]{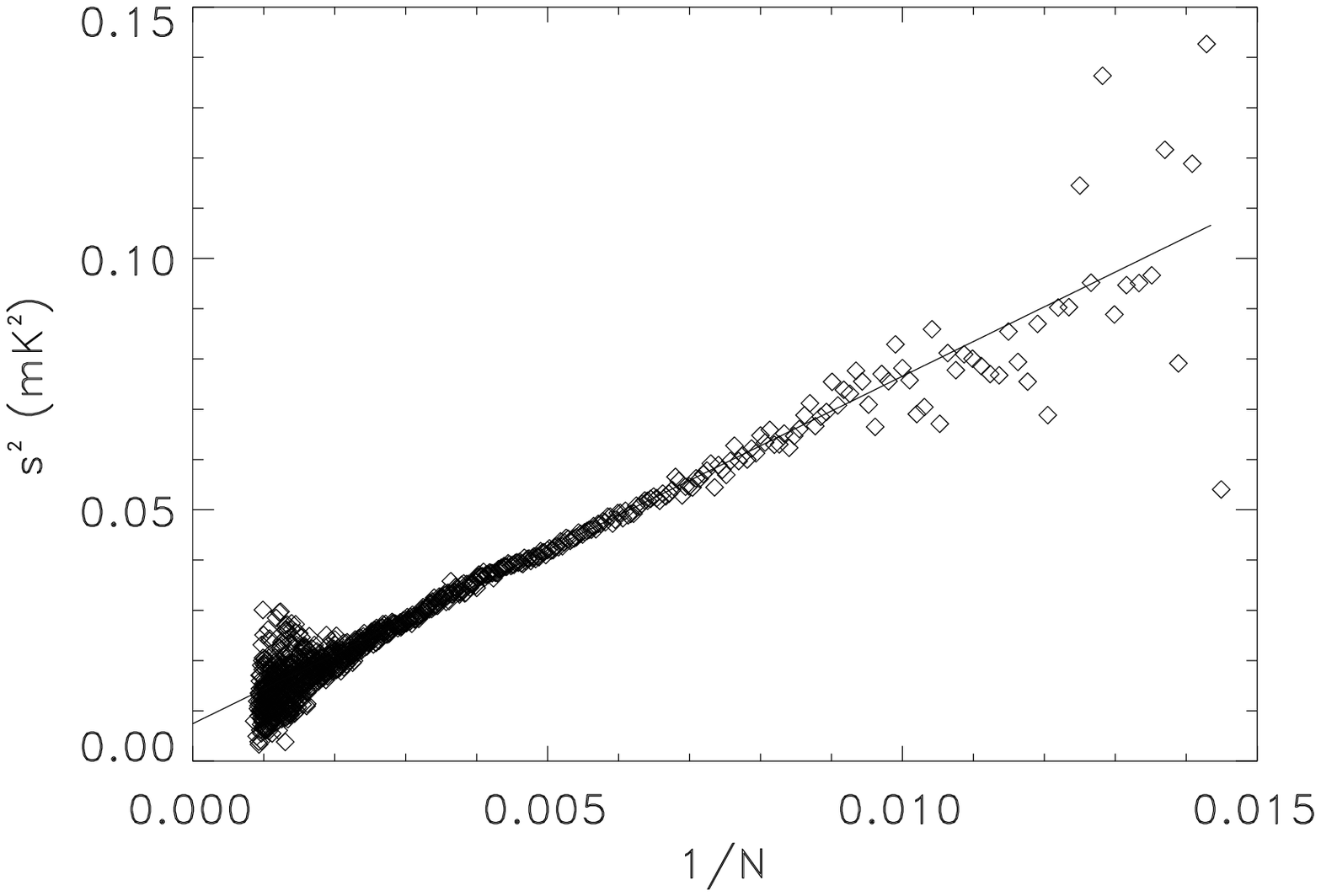}

\caption{ Measured sample variance, $s^2$,  dependence on normalized number of observations,  $N_{\rm obs}$,
 for the V1 sky map. The
diamonds are the sample variances  measured from each subset of  map
 elements for which $ N - 0.5 < N_{\rm obs} < N +0.5. $ 
The solid line is a linear fit to the data. As expected, the noise variance
scales as $N_{\rm obs}^{-1}$, confirming that it is a good predictor of the
pixel noise. }
\label{fig:noise_int}

\end{figure*}

Given the Gaussian noise distribution, the noise variance in a given pixel of the final sky maps should
scale inversely with the number of observations of  each pixel, $N_{\rm obs}$. 
This is tested by measuring the variance of each set of
map pixels with a given value of $N_{\rm obs}$ and plotting it 
against $N_{\rm obs}^{-1}$, as in Figure~\ref{fig:noise_int}.  Pixels with large Galactic
signals and point sources have been excluded based on the  Kp4 mask. (The values of $N_{\rm obs}$ have been
corrected to account for small variations in the radiometer noise levels resulting from changes
in the radiometers physical temperatures. Values of $N_{\rm obs}$ are therefore not integers, but are
scaled to correspond to a mean noise value.) 
Since the CMB fluctuations are uncorrelated with the instrument noise, they add a constant variance
to each bin. The slope of the line fit to the data is a measure of the noise per observation,
while the y-intercept is the variance arising from the CMB fluctuations. The noise variance
clearly scales as predicted, indicating that $N_{\rm obs}^{-1}$ is a good predictor of the
pixel noise.

\section{RADIOMETER GAIN MODEL}
The  primary gain calibration of \wmap~is derived from observations
of the CMB dipole. Although quite stable, small changes in
the radiometer thermal environment produce measurable changes in the radiometer gains.
Gain measurements with errors of $\approx 1\%$ are
obtained from each hour of observations. A gain model, relating
instrument housekeeping data to the observed radiometric gain, has been
developed to aid in smoothing and interpolating  the hourly gain measurements.
This model also reduces the uncertainty in gain determination when the dipole
derived gain measurements degrade as a result adverse  alignment of the scan pattern
with the dipole orientation or the Galactic plane. 

Following the notation of~\cite{jarosik/etal:2003}, the modulated microwave power
incident on a detector, $v_{\rm in}^2$, 
arising from  an antenna temperature difference, $\Delta T = (A^2 - B^2)$, may be written
as
\begin{equation}
v_{\rm in}^2 = \frac{1}{2}(A^2 - B^2)g_1 g_2 \cos(\theta). \label{egn:signal_1}
\end{equation}
 where $g_1$ and $g_2$ are the
voltage gains of the two amplification chains in the radiometer and
$\cos(\theta)$ is a term that accounts for phase mismatches in the radiometer.

\begin{figure*}
\includegraphics[width=6in]{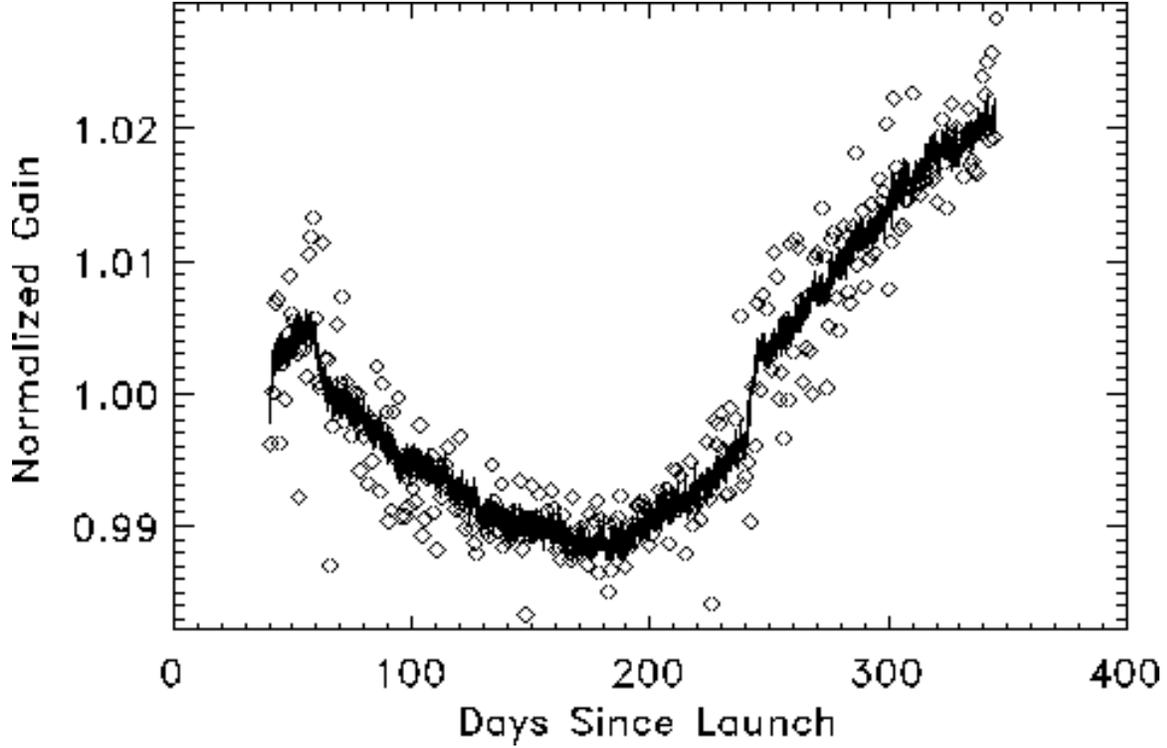}
\caption{ Comparison of the V11 radiometer gains measured from the CMB dipole (diamonds)
and the gain model (solid line). The CMB derived gains are 24 hour averages. The
high frequency noise in the gain model arises from the quantization of the 
RF bias housekeeping data. The rapid gain change $\approx 245$ days after
launch is the result of a small (0.5 K)  change of the RXB temperature as 
a consequence of an adjustment of the spacecraft's main bus voltage. Note the entire
vertical scale spans a $\approx 4\%$ range. The model clearly tracks
the measured gain values well and is  therefore useful in smoothing and interpolating
the hourly gain measurements.}
\label{fig:gain_model}
\end{figure*}

The output voltage from the detector with  input power $v_{\rm in}^2$ is
described by a response function $s(v_{\rm in}^2)$.
(This treatment differs from that given in~\cite{jarosik/etal:2003} in that the 
 response of the detector to the input  power is described by a non-linear response function,
$s(v_{\rm in}^2)$, to allow for deviations from square law response and 
gain compression which can occur in the warm HEMT amplifiers.)
 For the small ($< 2\%$) changes in detector input
power it suffices to linearize this response about the nominal operating point,
\begin{equation}
s(v_{\rm in}^2) = V_0 + \tilde{s} v_{\rm in}^2.
\end{equation}
where $\tilde{s}$ is the differential responsivity of the detector and $V_0$ is a parameter
to be fit.
The \emph{modulated} voltage output from the detector is then
\begin{equation}
S = \tilde{s}(v_{\rm in}^2) = \frac{\tilde{s}}{2}(A^2 - B^2)g_1 g_2 \cos(\theta). \label{eqn:mod_resp}
\end{equation}
The DC (unmodulated) power out of the same detector, $\overline{V}$, is given by
\begin{equation}
\overline{V} = V_0 + \frac{\tilde{s}}{2}\left\{\left(\frac{A^2 + B^2}{2} +  n_1^2\right) g_1^2 +
\left(\frac{A^2 + B^2}{2} +n_2^2\right) g_2^2\right\} \label{eqn:bias_1}
\end{equation}
where $n_1$ and $n_2$ are the instantaneous noise voltages added by the HEMT amplifiers
and the overbar indicates an average on time scales long compared to the period of
the microwave signal. The idea is to use the values of $\overline{V}$, measured to $\approx 0.1\%$
every 23.6 s, to estimate the value of $g_1 g_2$, which, according to equation~\ref{eqn:mod_resp},
determines the signal gain of the radiometer. Using the approximation
\begin{equation}
 \frac{(A^2 + B^2)}{2} + n_1^2 \approx  \frac{(A^2 + B^2)}{2} + n_2^2 \approx  n_0^2
\end{equation}
 Equation~\ref{eqn:bias_1} can
be rewritten as
\begin{equation}
 \frac{\overline{V} - V_0}{n_0^2} =  \tilde{s}~\frac{\overline{g_1^2} + \overline{g_2^2}}{2}
 \approx \tilde{s}\overline{g_1 g_2} \label{eqn:det_resp} 
\end{equation}
provided that
\begin{equation}
\frac{|g_1 - g_2|}{ |g_1 + g_2|} \ll 1 . 
\end{equation}
The term $n_0^2$ corresponds to the average system noise temperature, which
is dominated by the input voltage noise of the cryogenic HEMT amplifiers and as such
 is a slowly varying function of the physical temperature of the HEMT amplifiers, $T_{\rm FPA}$.
 Again, since the fractional
temperature variations are small, it suffices to describe small changes in noise
temperature using the relation
\begin{equation}
n_0^2(T_{\rm FPA}) = \alpha^{-1} (T_{\rm FPA} - T_0). \label{eqn:noise_dep}
\end{equation}
Combining equations~\ref{eqn:mod_resp}, \ref{eqn:det_resp}, and \ref{eqn:noise_dep} yields
\begin{equation}
S = \alpha \frac{\overline{V}-V_0}{T_{\rm FPA}-T_0}(A^2 - B^2).
\end{equation} 
Using measured values of the radiometer response, $S$, $T_{\rm FPA}$, and $\overline{V}$, and  the known differential dipole
signal, $A^2 - B^2$,  it is possible to determine the
parameters $V_0$, $T_0$ and $\alpha$. Once these parameters are determined the radiometer
gain can be estimated using the measured values of $T_{\rm FPA}$ and $\overline{V}$. 
Figure~\ref{fig:gain_model} shows  the
gain measured daily using the CMB dipole, and the result from fitting the gain model for the
V11 radiometer.
The model clearly tracks the dipole derived gain measurements well, and has substantially
less scatter. Values of the gain model parameters for all the detectors are presented in
\citet{limon/etal:2003}.

 Since the gain model depends only on the measured RF bias and temperature
of the FPA amplifiers, the fact that it fits the measured dipole gains well over such a long period
(1 year for this data set) indicates that radiometer characteristics are stable.
 Had some secular degradation occurred, for example a slowly increasing 
amplifier noise temperature or phase mismatch, constant values of the fit parameters,
$\alpha$, $T_0$ and $V_0$ would not provide a good fit to the measured  values for the entire
time period. Based on the accuracy of the fits there is no significant degradation in
performance of any of the radiometer channels at the 1\% level.

\section{SUMMARY}
All 20 differential radiometers aboard the \wmap~satellite are functional with
close to nominal performance. On-orbit values of the radiometer offset temperatures
are slightly higher than expected, leading to higher than expected $1/f$ knee frequencies
in two of the radiometers. Spin synchronous systematic errors from the radiometers 
and data collections system are  expected to be $< 0.17$ \ukelvin~for all radiometers, and the
radiometer noise distributions for all channels are well described by Gaussians.
 A gain model based on instrument
housekeeping data has been developed to aid in interpolating and smoothing the hourly gain calibrations
derived from the CMB dipole signal. No significant degradation in radiometer performance has been
observed during the first year of observations.  

\acknowledgements
 The \wmap~mission is made possible by the support of the Office of Space 
Sciences at NASA Headquarters and the efforts of numerous scientists, engineers,
machinists, managers, and office and administrative staffs at participating
institutions.

\acknowledgments

\end{document}